\documentclass[letterpaper, 10 pt, conference]{ieeeconf}  

\IEEEoverridecommandlockouts                              

\usepackage{graphics} 
\usepackage{subcaption}
\usepackage{graphicx}
\usepackage[utf8]{inputenc}
\usepackage[T1]{fontenc}
\usepackage{siunitx}[=v2]
\usepackage{hyperref}

\title{\LARGE \bf
Measurement of Individual Alteration in Perioperative ECGs During Elective Percutaneous Coronary Intervention
}

\author{Philip Gemke$^{a}$, Theresa Bender$^{a,b}$, Ennio Idrobo-Avila$^{a}$, Henning Dathe$^{a}$, Dagmar Krefting$^{a,b}$,
\\
Tim Kacprowski$^{c,d}$, Nicolai Spicher$^{a,b}$, \IEEEmembership{Member, IEEE} 
\thanks{$^{a}$Department of Medical Informatics,
        University Medical Center Göttingen, 37075 Göttingen, Germany
        {\tt\small philip.gemke@med.uni-goettingen.de}.}%
\thanks{$^{b}$DZHK (German Centre for Cardiovascular Research), partner site Göttingen, Göttingen, Germany.}     
\thanks{$^{c}$Division Data Science
in Biomedicine, Peter L. Reichertz Institute of TU Braunschweig and
Hannover Medical School,
        38106 Braunschweig, Germany.
        }
\thanks{$^{d}$Braunschweig Integrated Centre of Systems Biology (BRICS), TU Braunschweig, 38106 Braunschweig, Germany.}}

\def\ps@IEEEtitlepagestyle{%
  \def\@oddfoot{\mycopyrightnotice}%
  \def\@evenfoot{}%
}
\def\mycopyrightnotice{%
  {\footnotesize \hfill}
  \gdef\mycopyrightnotice{}
}

\begin{document}

\maketitle
\thispagestyle{empty}
\pagestyle{empty}


\begin{abstract}
The increasing availability of wearable electrocardiography (ECG) devices enables the continuous monitoring of individual ECG alterations. This could be beneficial for patients suffering from acute ischemia but with non-standard ECG findings that do not fit to the subject-independent and absolute thresholds defined in clinical guidelines. In this work, we evaluate the inter-patient magnitude of individual ECG alterations during ischemia.
The freely available STAFF III database provides 12-lead ECG recordings of patients before, during, and after elective percutaneous coronary intervention (PCI), where a coronary vessel is widened with a balloon inflation. We compute individual alterations of ST-interval and T-wave amplitudes w.r.t. QRS amplitude over time for each patient and lead. We demonstrate that determining relative ST-interval/T-wave amplitudes and deriving individual alterations over time is feasible in standard and non-standard ECG recordings. To demonstrate clinical relevance, we use the features for differentiating N=$\mathbf{54}$ STAFF III patients with atherosclerotic plaque in either the right coronary artery (RCA) or left ascending artery (LAD). Results show significant differences in $\mathbf{5}$ leads for ST-interval alterations and $\mathbf{3}$ leads for T-wave alterations, which are also suggested by clinical guidelines for ischemia detection. 
\textit{Clinical relevance}— Assessing individual alterations over time could eventually close the gap in ECG evaluation of patients presenting with pre-existing heart conditions and non-standard ECGs. 
\end{abstract}

\section{INTRODUCTION}
Electrocardiograms (ECGs) are recordings of the electrical activity of the heart, frequently showing abnormalities that can be indicators of diseases. Abnormalities are usually assessed based on subject-independent and absolute thresholds given in clinical guidelines~\cite{c1}. 

However, individual alterations below defined thresholds can be missed during manual ECG interpretation. In the past, these individual alterations were hard to take into account, since previous ECG recordings were on paper in analog medical records and thus frequently unavailable for the treating clinician. The upcoming availability of digitized ECG recordings and easy-to-use wearable devices blazes the trail for individualized evaluation of alterations w.r.t. previously-recorded ECGs~\cite{c2}. 

In an elective PCI the chronically narrowed coronary arteries are opened to enable sufficient oxygen supply of the heart itself and improve quality of life for patients suffering from coronary artery diseases (CAD).
In this non-surgical technique a catheter is inserted into an artery and advanced to the heart. After the exact location of the narrowed coronary vessel is determined, a balloon is inflated to widen the artery. In early days of PCI, the focus was prolonged balloon inflation time in order to compress the atherosclerotic plaque with high pressure against the arterial wall, resulting in improved blood flow~\cite{c4}. 

The ECGs recorded during this outdated technique imitate the ECG alterations resulting from blocked coronary vessels, since there is strongly reduced blood flow in the vessel during balloon inflation. In the last decades, brief inflations with lower pressure and the usage of metal or plastic stents have been established to further reduce risk of life threatening complications~\cite{c3}. For example, restenosis which is a new occlusion of the treated localization after PCI.
Patients undergoing elective PCI have pre-existing heart conditions and often show non-standard ECG recordings with permanently altered amplitude of the ST-interval and T-wave, which is hard to interpret using general guidelines~\cite{c14, c15}. Thus, we investigate relative ST-interval/T-wave amplitudes and derive individual alterations over time.

As a clinical use case, we differentiate patients undergoing PCI in either the RCA or LAD. The first supplies the posterior right side of the heart while the latter supplies the left anterior side of the heart and often shows laterally reversed ECG findings. We use the ST-interval amplitude as the slowed electric impulse, pointing in the direction of the ischemic region, can be detected on the affected site as ST-elevation and on the non-affected side as ST-depression~\cite{c6}. Since the location of the narrowed coronary vessel correlates with the leads in which an alteration is visible, this allows to pinpoint occlusions using the ECG. 
\vspace{1.5cm}
\\
\mycopyrightnotice{\textit{\textsuperscript{\textcopyright} 2023 IEEE.  Personal use of this material is permitted. Permission from IEEE must be obtained for all other uses, in any current or future media, including reprinting/republishing this material for advertising or promotional purposes, creating new collective works, for resale or redistribution to servers or lists, or reuse of any copyrighted component of this work in other works.}}

\section{METHODS}
We aim for quantifying individual ECG alterations in ischemic conditions. Thereby, we selected data of patients undergoing PCI and focus on the amplitude of the ST-interval, where regularly no electric activity of the heart can be measured in physiologic ECG recordings, and the amplitude of the T-wave, representing the repolarization of the ventricles. Both features are based on clinical guidelines~\cite{c1} for diagnosis of ischemia with the amplitude of the ST-interval being an established parameter~\cite{c6,c9,c12}.

\subsection{Dataset}
The STAFF III database contains 12-lead ECGs from $104$ patients treated with the outdated technique of elective prolonged balloon inflation in a major coronary artery~\cite{c5}. The ECGs were acquired from 1995 to 1996 at Charleston Area Medical Center (WV, USA) and sampled at \SI{1000}{\hertz} with an amplitude resolution of \SI{0.6}{\micro\volt}. Custom-made equipment by Siemens-Elema AB (Sweden) was used to record standard chest leads and limb leads in the Mason-Likar configuration~\cite{papouchado1987fundamental}.

The pre- and post-inflation ECGs were both recorded \SI{5}{\minute} in supine position at rest, in a baseline room before and lying in the catheter lab after PCI. The mean inflation time was \SI{4}{\minute} \SI{23}{\second}, ranging from \SI{1}{\minute} \SI{30}{\second}  to \SI{9}{\minute} \SI{54}{\second} with exact annotations provided, including time and localization of the balloon inflation~\cite{c5}.  

All patients that underwent primary balloon inflation in the RCA or LAD were processed. We analyze $23$ patients (age: $60 \pm 12$ years; $48\%$ female) with balloon inflation in the LAD and $31$ patients (age: $58 \pm 8$ years; $42\%$ female) with balloon inflation in the RCA. For each patient, several ECGs were measured at three different stages: i) pre-inflation, ii) during-inflation and iii) post-inflation.

\subsection{Preprocessing}
In order to reduce noise, a single representative beat~\cite{c11} is calculated using the function \texttt{ecg\_segment} of the open-source library Neurokit2~\cite{c7}. For this computation, single heart beats are temporally aligned using the R-peaks detected by Kalidas algorithm~\cite{c8}. Additionally, the quality of these heart beats is calculated using the function \texttt{ecg\_quality} of Neurokit2~\cite{c7} and only heart beats with at least \SI{80}{\%} similarity to the others are included. After selecting the lead with the highest R-peak amplitude, the ECG signal is delineated using the function \texttt{ecg\_peaks} of Neurokit2~\cite{c7}. The T-, P- and R-peaks of this lead are then transferred to all other leads and their positions are corrected within a range of \SI{40}{\milli\second} to address minimal deviations between leads. 

\subsection{Feature engineering}
We define the ST-interval as a \SI{40}{\milli\second} segment starting \SI{85}{\milli\second} after the R-peak according to García \textit{et al.}~\cite{c9}. The isoelectric line is defined as the mean amplitude of the TP-interval, calculated within physiologic boundaries of \SI{50}{\milli\second} after the T-peak and \SI{40}{\milli\second} before the P-peak. This enables to evaluate the amplitudes of the ST-interval and the T-peak in relation to QRS-amplitude, which we will denote ''ST feature'' and ''T feature'', respectively. According to Aslanger \textit{et al.}, this approach based on relative differences has the advantage of including the cardiac vector information in contrast to absolute quantification in \SI{}{\milli\volt}~\cite{c6}.

For each patient and lead, features are computed in overlapping intervals of \SI{50}{\second} with steps of \SI{10}{\second}. In order to quantify the individual alterations of the features measured during stages pre-inflation (i) and during-inflation (ii), the difference between average values of both stages is computed (ii-i), which we will denote ''$\Delta$ST'' and ''$\Delta$T''.

\subsection{Statistical analysis}
In order to evaluate the differences in individual alterations between LAD and RCA patients, we test for statistical differences. First, the null hypotheses whether the groups are normally distributed is tested in every lead using a Jarque–Bera test. Second, the null hypothesis is tested that the variances of both groups do not differ for every lead and feature using a two-sided t-test.

\section{Results}

\begin{figure}[t!]
    \centering
    \framebox{\parbox{3.3in}{ \includegraphics[trim={2cm 2cm 3cm 4cm},clip, scale=0.24]{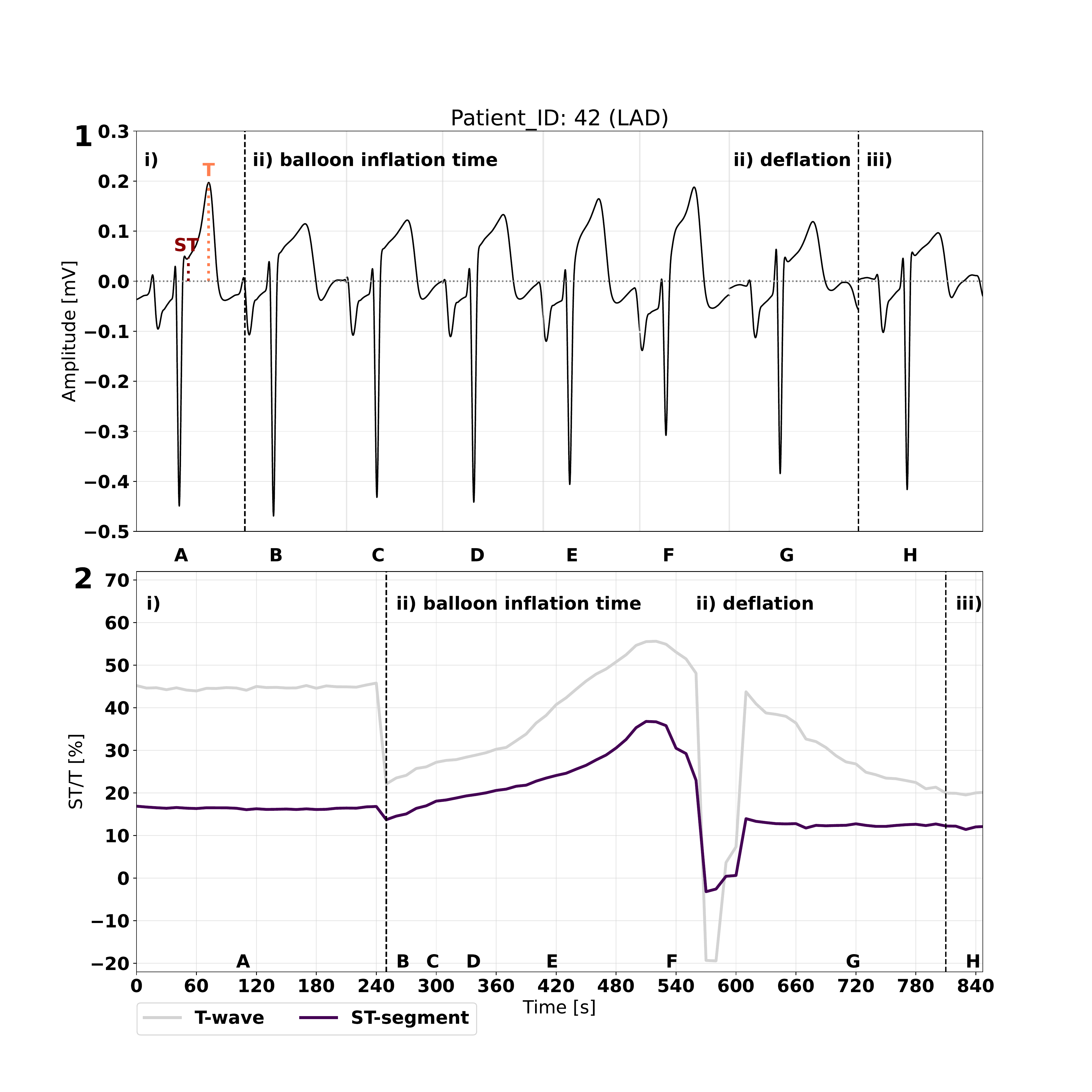}
    
    \vspace{.5cm}
  
    \includegraphics[trim={2cm 2cm 3cm 4cm},clip,scale=0.24]{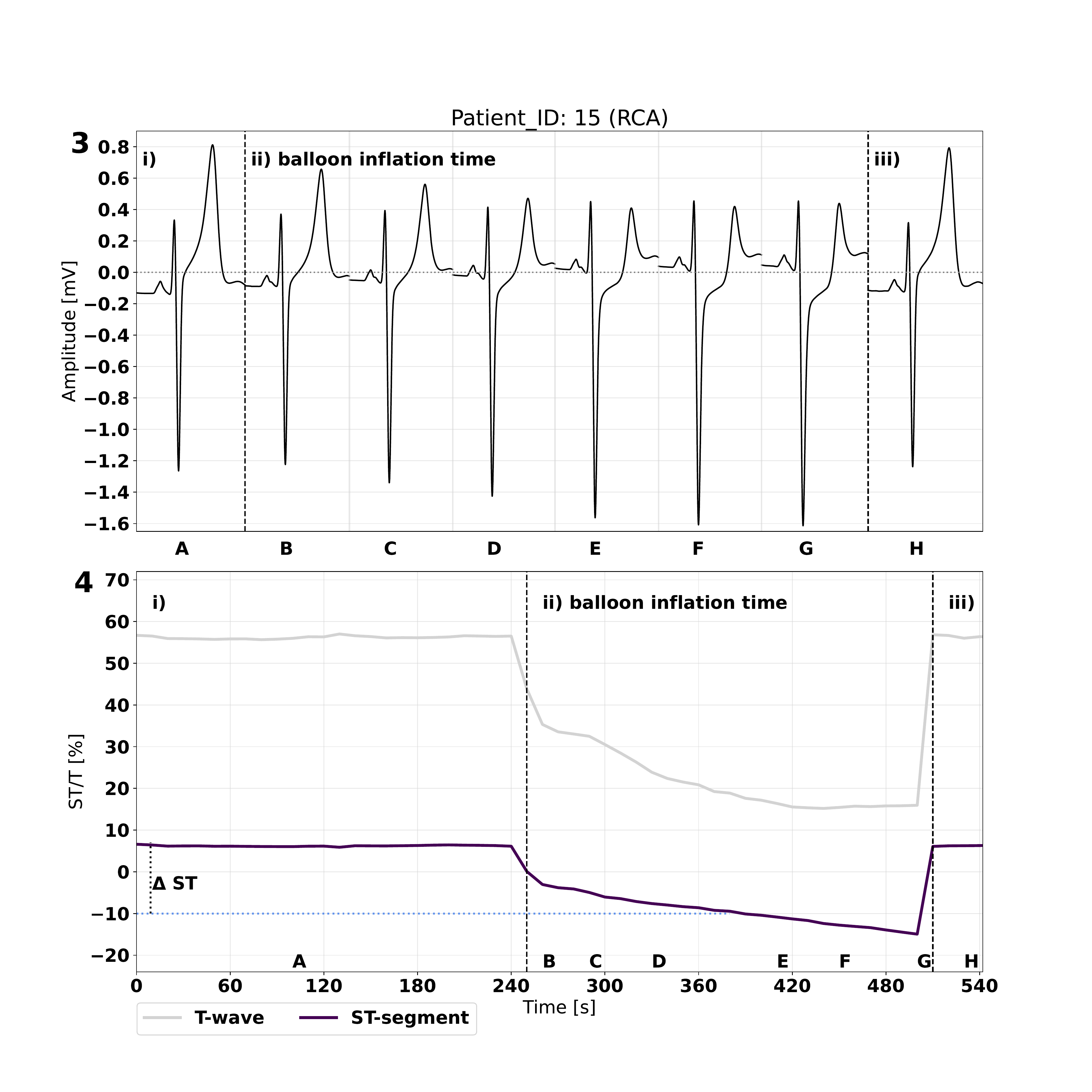}
}}
    \caption{Lead V2 features during elective PCI for an RCA patient (panels 1-2) and LAD patient (3-4). Representative beats (1,3) and ST and T feature (2,4) computed over time during three stages are shown: i) pre-inflation ii) during-inflation iii) post-inflation. In the LAD patient, stage ii) additionally shows the moment of balloon deflation within the recording. Vertical, dashed lines mark interruptions in measurements due to medical intervention. A-H show occurrences in time of the representative beats and their value w.r.t. QRS amplitude. In panel 1, the ST and T feature of the first representative beat are shown. In panel 4 bottom left, the computation of the individual alterations $\Delta$ST are depicted exemplary.}
    \label{fig:examples}
\end{figure}

Fig.~\ref{fig:examples} shows representative beats (panel 1,3) and the individual alterations (panels 2,4) for an RCA and LAD patient, respectively. 
Lead V2 shows individual alterations for ST mainly within the balloon inflation time. During occlusion of RCA, this results in a constant decrease of ST and T features, visible in the representative beat (panel 1) and the calculated relative values (panel 2). In contrast, an occlusion of LAD results in a constant increase of ST and T within the representative beat (panel 3) and relative values (panel 4). The LAD patient shows a non-standard ECG, presumably due to a pre-damaged heart, which is expressed by ST-elevation before balloon inflation. Regarding the T feature, individual alterations (panels 2, 4) show a similar behavior in the during-inflation stage.

\begin{figure}[]
    \centering
    \framebox{\parbox{3.3in}{ \includegraphics[scale=0.24,trim={2cm 2cm 3cm 4cm},clip,]{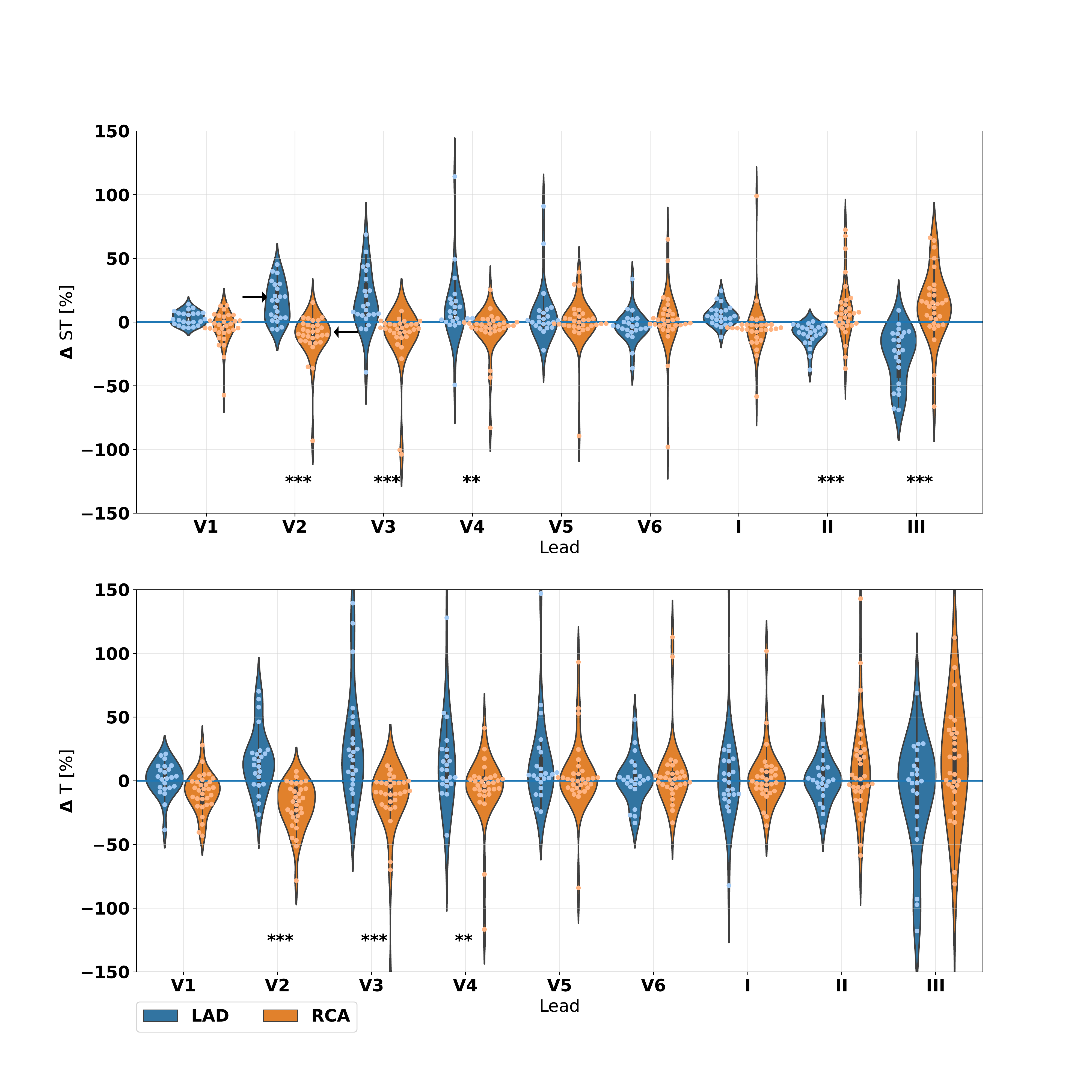}
}}
    \caption{$\Delta$ST and $\Delta$T features of LAD (blue) and RCA (orange) patients. Each point corresponds to a patient. In leads V2-V4 both groups can be easily distinguished from each other. The black arrows indicate the data points for patients 42 and 15 in $\Delta$ST. $\ast\!\ast\!\ast$ indicates a significance with $p \leq 0.001$ and $\ast\ast$ indicates $0.001< p \leq 0.01$.}
    \label{fig:Violine}
\end{figure}

Fig.~\ref{fig:Violine} summarizes $\Delta$ST and $\Delta$T features representing the differences between the stages pre-inflation and during-inflation in every lead and patient. For $\Delta$ST, leads V2-V4 show higher values for LAD patients compared to RCA patients. In leads II-III the results are reversed with RCA patients showing higher values.
Regarding $\Delta$T, in leads V2-V4 the same effect as in $\Delta$ST can be seen, while leads II-III do not show clearly effects.
The differences in leads V2-V3 and II-III are highly significant for $\Delta$ST, whereas in $\Delta$T only V2-V3 are highly significant.

\section{Discussion}

The proposed individual ECG alterations during ischemia of patients undergoing prolonged balloon inflation show a potential for differentiating LAD and RCA patients. The features in the LAD patients are very similar to clinical findings, where standard ECGs show characteristic ST-elevation in leads V2-V4 with contralateral ST-depressions in leads II and III~\cite{c6}. \textit{Vice versa}, laterally-reversed ECG findings can be found in RCA patients using our features which is also covered by clinical guidelines~\cite{c6}. 

However, clinical guidelines~\cite{c1} use 10-second 12-lead ECGs representing only a snapshot of the current cardiac activity and look for absolute alterations in \SI{}{\milli \volt}. We demonstrate that it is also feasible to detect individual alterations over time based on relative values~\cite{c6}. In Fig.~\ref{fig:examples} the pre-inflation ECG of LAD patient 42 already shows an elevated ST-interval amplitude. This ST-elevation is problematic as the conventional assessment prevents proper diagnosis of life-threatening restenosis. However, our proposed individual alterations $\Delta$ST and $\Delta$T show characteristic changes over time, which could be an avenue for future work for more accurate diagnosis. 

In the recent guideline for evaluation of chest pain~\cite{c1}, the T-wave is not taken into account considerably. However, there are many researchers trying to improve diagnosis based on T-wave features~\cite{c12}. Within the STAFF III database, García et al.~\cite{c9} found that in $72\%$ of patients with ST-interval amplitude changes were also showing T-wave amplitude changes. Martínez et al.~\cite{c10} studied the prevalence, magnitude and spatio-temporal relationship between T-wave alternans and ischemia. They analyzed the ECGs beat by beat and found T-wave alterations in $33.7\%$ of the patients. A reason why usage of T-waves is challenging can be seen in the pre-inflation ECG of patient 42 (panel 1) in Fig.~\ref{fig:examples}: The T-wave (A) shows a similar height as during balloon inflation (F), highlighting its morphological variability. A potential explanation could be that the ECG was detached and re-attached between both stages. 

However, our results indicate that the T-wave might be a suitable additional feature for measurement of ischemic ECG alterations if the ECG device is constantly attached. Additionally, our T feature shows a similar behavior in leads V2-V4 during balloon inflation compared to the ST feature. In contrast, this cannot be observed in leads II-III. This could be explained by the fact that leads V1-V4 are placed directly on the chest wall and result in more pronounced signal amplitudes.

A limitation of this work is that we only analyzed a single database. Therefore, investigating whether the observed alterations can also be noticed in other patient cohorts and databases is an avenue for future work. Furthermore, our algorithms are based on computing a representative beat which potentially is not able to capture fine details. Beat-to-beat approaches~\cite{c12} could be more sensitive for detecting the dynamics of ischemic alterations in the first minutes after occlusion of a coronary vessel. 

\section{CONCLUSIONS}
In this work, two established ECG features, namely the amplitudes of the ST-interval and T-wave~\cite{c1, c2, c3}, were automatically extracted and quantified over time in their relative distance to the QRS complex amplitude. We found significant differences between LAD and RCA patients when comparing ECGs before and during PCI. This approach extends the state-of-the-art based on subject-independent and absolute thresholds defined in clinical guidelines~\cite{c1}. The individual features could be useful for patients in which these “one size fits all” thresholds fail, e.g. in patients showing only subtle alterations~\cite{jansen2021chest} or with permanent ST-alterations~\cite{manne2018atrial}.

The proposed method could eventually prepare the ground for wearable devices in long-time monitoring. As the features give intuitive and physiologically-interpretable insight into the cardiac state, their visualization could be helpful for rapid assessment of long-time measurements.

\addtolength{\textheight}{-12cm}   




\bibliographystyle{ieeetr} 
\bibliography{lib}

\end{document}